**Supercurrent-Controlled Kinetic Inductance Superconducting Memory**

Eduard Ilin, Xiangyu Song, Irina Burkova, Andrew Silge, Ziang Guo, Konstantin Ilin, and Alexey Bezryadin

Department of Physics, University of Illinois at Urbana-Champaign, Urbana, IL 61801, USA

**Abstract**: We report superconducting kinetic inductance memory (SKIM) element, which can be controlled exclusively by the bias supercurrent, without involving magnetic fields and heating elements. The SKIM is non-volatile memory. The device is made of Nb and it can operate reliable up to 2.8 K. The achieved error rate is as low as one in 100000 operations.

**Keywords**: superconducting memory, nanowire, kinetic inductance, phase slip, fluxon, fluxoid.

## 1. Introduction

Superconducting supercomputers could provide a powerful alternative to modern semiconductor-based computers since the power dissipated in superconducting elements can be very low, i.e., it can approach thermodynamic and quantum of the information processing [1,2,3,4,5,6]. The "bottleneck" for the development of superconducting computers is the absence of a compact fully superconducting memory element. Nanowires [1,4,7,8,9] provide an important alternative to the existing elements based on tunnel junctions [10,11,12,13,14,15,16,17] since their operation is based on their kinetic inductance [1,3,18] which could be very large in nanowires with nanometer-scale dimensions. The employment of a large kinetic inductance for the memory function allows for a great reduction of the dimensions of the device [1,9].

Previously reported superconducting memory either required magnetic fields for its functioning [1] or was not fully superconducting since some of its elements required Joule heating [2]. Here we report a fully superconducting kinetic inductance memory (SKIM) element, based on single quantized superconducting fluxons, which is free from previously existing drawbacks listed above. The SKIM is made of Nb, deposited over a pair of suspended SiN nanoconstrictions. The reported memory does not require sub-Kelvin temperatures, as was the case in many previously reported devices. Our SKIM element was tested in a cryogen free refrigerator and was operational up to a temperature of ~2.8 K. The highest frequency of operation, tested up to ~50 kHz, was limited by the setup equipment and could be much higher since the phase slip rate in nanowires and nanoconstrictions is of the order of 100 GHz [19].

## 2. Sample fabrication

The sample geometry is shown in Fig.1a. It consists of two nanobridges connected at each their end to thin superconducting films, the "electrodes". The electrodes and the nanoconstrictions are made of 15 nm thick Nb film. To fabricate the sample, a line-with-break pattern is prepared by e-beam lithography with line dosage 9 nC/cm, on a $SiN/SiO_2/Si$ substrate. The trench and nanobridges are developed in a 2 min 30 s SF6 reactive ion etching (RIE) procedure and a 10 s HF wet etching, needed to form an undercut, to make the nanobridges suspended. The undercut is visible in Fig.1a as a gray area on both sides of the trench. The resulting structure contains a 200 nm wide trench (dark area in Fig.1a) and two constrictions bridging the gap over the trench. The width of the SiN nanobridges is typically in the range 10-30 nm. A 15 nm Nb film is sputtered over the SiN nanobridges to form superconducting nanoconstrictions, the width of which is defined by the width of the underlying SiN bridge. Initially there are many nonobridges crossing the trench. We use photolithography to select and protect a pair of nearby nanoconstrictions and to define the electrodes. The unprotected Nb is removed by 2 min 30 s SF6 RIE.

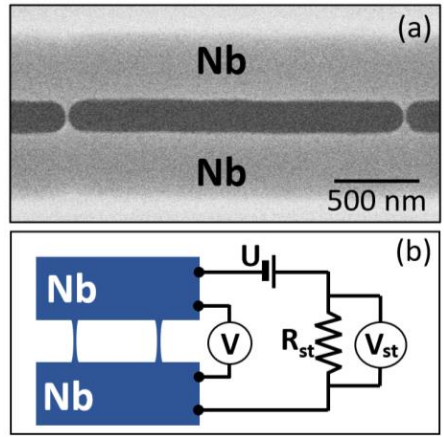

*Figure 1. (a) Scanning electron micrograph of the sample. Two Nb nanobridges connect the top Nb electrode and the bottom Nb electrode. The Nb electrodes are separated by a 200 nm wide trench (black). The Nb nanobridges are suspended over the trench. (b) Schematic of the memory element. The element contains a loop formed by the two bridges linked to two Nb electrodes. Such superconducting loop can have different vorticity states and serves as a non-volatile memory element, able to store one bit of information.*

## 3. Experimental details

The sample was placed in a cryogen-free cooler RDK-101DL (Sumitomo Heavy Industries, Ltd.) providing the base temperature of 2 K. The leads were thermalized by passing them through copper powder RF radiation filters, filled with a fixture of Stycast epoxy and Cu particles [20]. The current bias of the sample was set by taking an ac voltage, $U$, from a high precision source DS360 and applying it to the sample connected in series with a standard resistor of $R_{st}$=1 kΩ (see Fig.1b). The sample resistance was much lower than the standard resistor. The typical frequency was 11 Hz, low enough so that further reduction of the frequency would not change the results. The voltages on the sample, $V$, and on the resistor, $V_{st}$, were measured using National Instrument data acquisition card NI-DAQ USB-6216. The current in the circuit, $I$, was calculated as $I=V_{st}/R_{st}$. The voltage taken from the sample and the resistor was amplified using low noise differential preamplifier SR560 Stanford Research Systems, before being supplied to the DAQ card.

## 4. Results

The temperature dependence of the resistance ($R=V/I$) of the sample is shown in Fig.2a. There are two curves on the graphs, one is taken at low bias current (the red curve), which does not impact the shape of the curve significantly, while the other (blue) is taken at a much higher current, which was smaller but of the same order as the critical current of the nanowires. The first curve shows the true temperature dependence and confirms that nanobridges have the same critical temperature as the film. But since they have a much lower critical current (since they are much narrower than the film) the curve taken at a higher bias current shows two transitions. Such R-T curve shows that the normal resistance of the film was ~85 Ohm and the normal resistance of the nanobridges was ~15 Ohm. Taken the dimensions of the nanobridges (20 nm by 15 nm) we estimate the resistivity 9 µΩ cm, which is a typical value for Nb thin film [21].

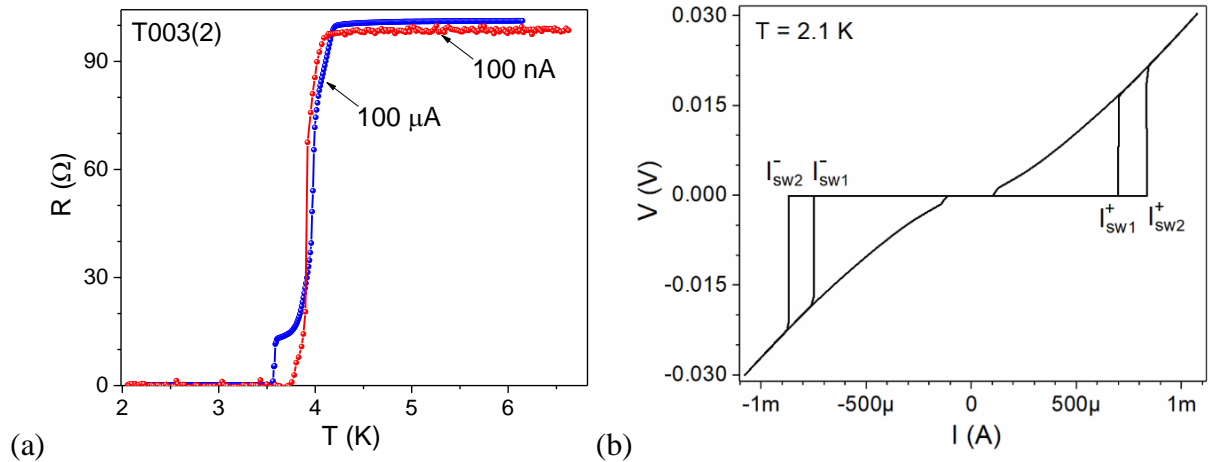

Figure 2. The sample resistance plotted versus the temperature, T. There are two curves, one (red) measured at a very low bias current (100 nA) and the other (blue) taken at a higher bias current (100 µA). The critical temperature is $T_C = 4$ K. (b) The I-V-curves are measured at $T = 2.1$ K.

The voltage versus current curve ("V-I curve") is shown in Fig. 2b. It exhibits two different switching currents, of the order of 1 mA. The origin of the two switching currents, as will be discussed later, is due to different possible values of the winding number of the superconducting condensate wave function defined on the loop formed by the two nanobridges. This winding number will be used to store information. It should be emphasized that if the curve would be measured just one time then it would show just one value of the critical current. Yet if the bias current is cycled many times, as is the case in Fig. 2b, then the multivalued V-I curve, having two switching currents, emerges. The V-I curve also shows hysteresis, and the re-trapping current appears much lower than the switching current. We explain this by Joule heating. Note that the voltage observed after the switching is of the order of 15 mV, which is in agreement with the normal resistance of the nanobridges estimated from the R-T-curves (see above).

We have also tested the temperature dependence of the critical currents ("switching currents") (Fig.3a). It appears that the distinction between the two critical currents becomes smaller and almost disappears at ~3.2 K. Since the distinction between the critical currents will be used to determine the winding number of the loop, the temperature has to be substantially lower than ~3.2 K for the memory to operate reliably.

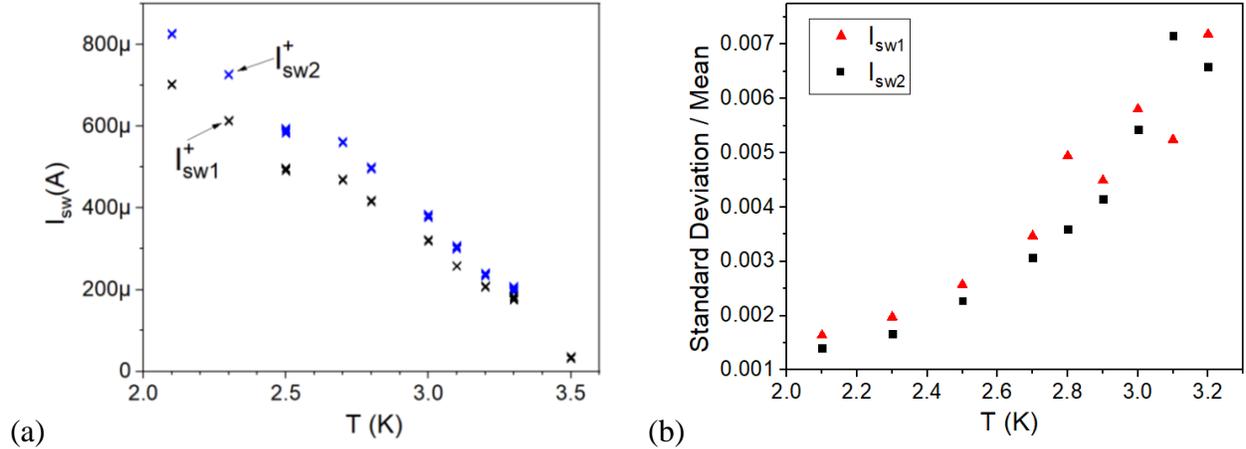

*Figure 3. (a) The two possible values (black and blue) of the critical current ("switching current") plotted versus the temperature. (b) Dispersion of the switching current. The standard deviation is computed over an ensemble of 10000 points. It is normalized by the mean value of the switching current.*

The fluctuations of the switching current have also been measured (Fig.3b). It is clear that the standard deviation is of the order of 1 µA, which is much smaller than the difference between the two critical currents (~0.1 mA). Thus, these fluctuations cannot cause the memory switch between the states.

Figure 4 shows the dependence of the switching current on the perpendicular magnetic field. The function is periodic and multivalued. Each branch of this periodic function represents a certain vorticity of the memory loop, or, in other words, the winding number of the order parameter, or, in other words, the number of fluxoids trapped in the loop [9]. This number will be denoted as $n_v$. The observed periodicity shows that vortices do not sit in the wires themselves but enter the space between the wire. Also, it shows that the vortices enter the loop in a periodic manner, as the magnetic field is increased. The key observation is that the function is multivalued even at zero field. Thus, we conclude that two vorticity states are possible at zero field. These states can be used to store one bit of information. Below we will discuss how to switch the system between the two stable vorticity states by the bias current, even as the magnetic field remains zero at all times. Generally speaking, this will be possible because a state with a fixed number of vortices in the loop ($n_v$) has different positive and negative critical currents, due to the fact that the bridges are not identical (asymmetry).

The reason that the switching current has two possible values at zero magnetic field is that the system can have vorticity zero ($n_v = 0$) or a unit vorticity ($n_v = 1$) when a vortex is trapped in the loop [10]. As the bias current is applied, it is added to the persistent current in the loop. So, the observed switching current is lower in the case $n_v = 1$.

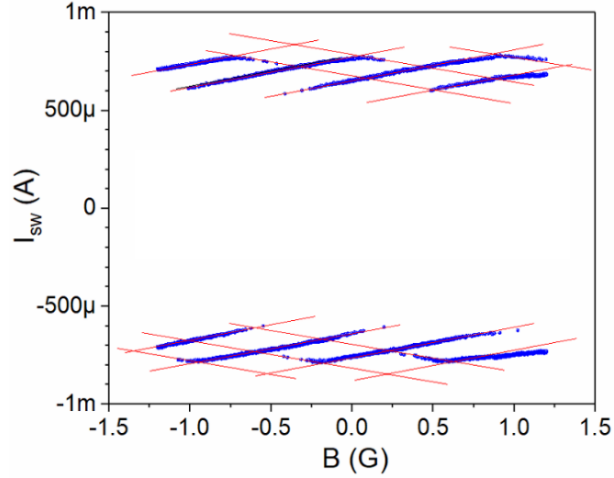

*Figure 4. Magnetic field dependence of the switching current (critical current). The multivalued nature of the switching current is clearly visible, representing two different vorticity states of the loop, formed by the pair of parallel Nb nanobridges.*

The main question is how to switch the system between the states without applying external magnetic field. Our method is as follows. We apply a control current $I_0=(I^+_{sw1} + I^+_{sw2})/2$ and measure the voltage. If the voltage is zero that means the $n_v=0$. This memory states are considered logical zero (mem=**0**). If the voltage is not zero, then the bias current is reduced to zero and increased again to the value $I_0$. Again, if the voltage is zero then the logical zero state is considered created (mem=**0**). This cycle is repeated until the voltage is zero when the current is $I_0$ applied. After a few cycles, the state mem=**0** is always created with this approach.

To write the state mem=**1** a similar procedure is used, but the applied current is negative, namely $I_1=(I^-_{sw1} + I^-_{sw2})/2$. Note that here the switching currents are negative $I^-_{sw1}<0$ and $I^-_{sw2}<0$.

To read out the state of the memory we always apply the current $I_0=(I^+_{sw1} + I^+_{sw2})/2$, and measure the voltage on the device. If the measured voltage is zero, then the state was mem=0; if the voltage is a few mV, then the state was mem=**1**, before the measurement. The state is destroyed if mem=**1** is measured, since in this case the system switches into the normal state and the order parameter is destroyed. By repeating the critical current measurement many times, we can plot distributions as is shown in Fig.5. The top plot (Fig.5a) was measured without preparing the sample state anyhow. Thus, a random sequence of the critical currents $I^+_{sw1}$ and $I^+_{sw2}$ was measured. The corresponding distribution (black) shows two peaks due to the fact that the sample can be found in two different vorticity states, $n_v$. The reason we find different switching currents is that as the bias current is changing from positive to negative periodically.

Next, we perform a test in which the vorticity state is set to $n_v=1$ (mem=**1**) before each measurement of the switching current. In this state the system shows the lower critical current $I^+_{sw1}$ at the positive bias, and the higher critical current, $I^-_{sw2}$, at the negative bias. Note that $|I^-_{sw2}|>|I^+_{sw1}|$.). The corresponding distribution is shown by the red bars in Fig.5b, measured at positive bias currents. The distribution shown by the blue bars (Fig.5c) was measured in a similar way, but the state of the memory was mem=**0** ($n_v=1$) before each measurement of the switching current. The key observation from Fig.5 is that it is possible to select the desired state of the memory and the distribution becomes a well-defined single peak, with only one error out of about 200000 measurements.

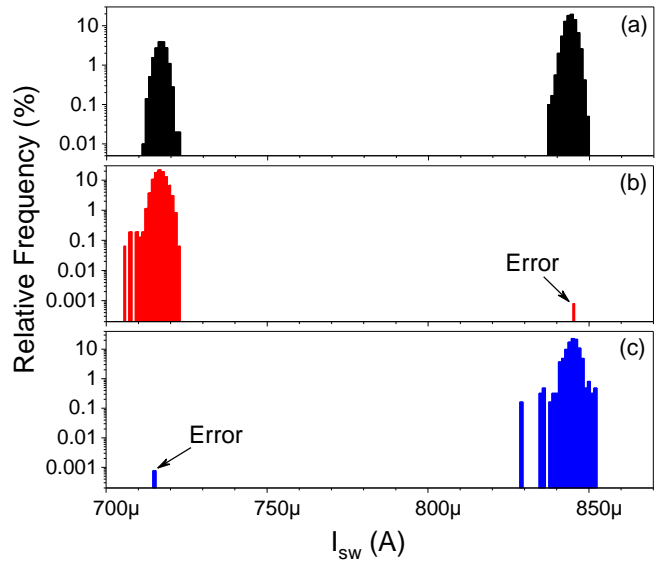

*Figure 5. Distribution of the switching current: (a) Switching current of the unconditioned sample. (b) The sample was conditioned to mem=**1**, before each switching current measurement. (c) The memory element was set to mem=**0** before each measurement. T = 2.1 K. The total number of measurements used to plot each one of these three distributions was about 200000.*

The measurement of Fig.6a illustrates the operation of the memory. Here we write a state and then read out the state and verify that the state read out is the same one as the state created before the measurement. The sequence of written states was 0000-1111-0000-1111 etc. The sequence of read out states is exactly the same, as is shown in Fig.6a. Within this example, a total of 28 write-read cycles is performed and no errors observed. If the test is continued for 100000 write-read cycles, then we find a few errors. Thus, the error rate is of the order of $10^{-5}$.

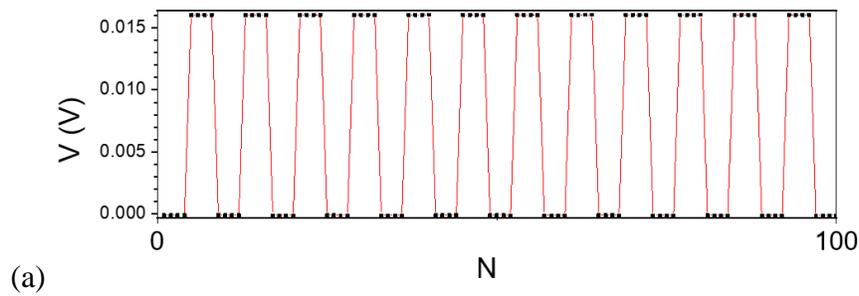

(a)

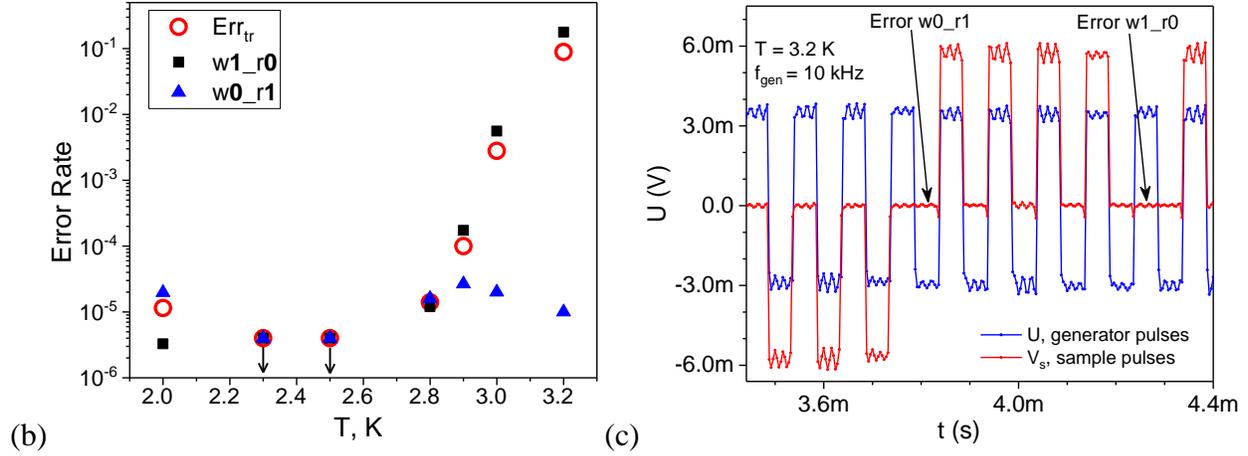

*Figure 6. (a) In this test run we write mem=0 state four times. After each writing we read out the state and put a point on the plot, which is shown as the voltage on the device. Then we write the mem=1 state, also four times. After each such writing the state is read out and the resulting voltage on the device is put as a point on the plot. Thus, we observe four mem=0 states then four mem=1 states then four mem=0 states etc. No errors are observed. (b) Various error rates, w0_r1, w1_r0, and Err$_{tr}$ = w0_r1+ w1_r0 are plotted versus the device temperatures. The data points having the arrows attached to them provide the top limit estimate for the error rate. The exact values are unknown since within N~300000 measurements no errors have been detected. (c) Definition of an error in a higher frequency reading process.*

The two types of error rates (write-0-read-1 error rate, w0_r1, and write-1-read-0 error rate, w1_r0) were calculated by dividing the error number by the number of measurements (Table 1, Fig. 6b). The total error rate, $Err_{tr}$, is calculated by dividing the sum of the two types of errors by the total number of measurements, $N$, which was typically $N$~300000.

*Table 1. Error rates for various temperatures.*

| Error Rate | $T$=2K | $T$=2.3K | $T$=2.5K | $T$=2.8K | $T$=2.9K | $T$=3K | $T$=3.2K |
|---|---|---|---|---|---|---|---|
| $Err_{tr}$ | $1.1 \cdot 10^{-5}$ | 0 | 0 | $1.4 \cdot 10^{-5}$ | $1 \cdot 10^{-4}$ | $2.8 \cdot 10^{-3}$ | $8.8 \cdot 10^{-2}$ |
| w0_r1 | $1.9 \cdot 10^{-5}$ | 0 | 0 | $1.6 \cdot 10^{-5}$ | $2.6 \cdot 10^{-5}$ | $2 \cdot 10^{-5}$ | $1 \cdot 10^{-5}$ |
| w1_r0 | $3.2 \cdot 10^{-6}$ | 0 | 0 | $1.2 \cdot 10^{-5}$ | $1.7 \cdot 10^{-5}$ | $5.5 \cdot 10^{-3}$ | 0.18 |

At higher frequencies (~10 kHz) the error rates have been determined by applying a sequence of current-bias square pulses, switching between the positive limit $I_0$ and then to the negative limit $I_1 < 0$. The result is shown in Fig.6c. In this case we show the voltage on the sample (red) and the voltage on the standard resistor (blue) versus time. The current is proportional to the voltage on the standard resistor. If the current is positive and the sample voltage is zero, the state is mem=**0**. Then the current switches to the negative bias. If the voltage is high that means we read out mem=**0**, so, there is no error. If, on the other hand, the voltage remains zero even when the current changes its polarity to negative, then the state is mem=**1**, so we count this event as an error. Thus, we test the error rate w0_r1. Analogously, the other error rate, w1_r0, was tested also. The resulting total error rate was 2.3·10$^{-5}$ for the 10 kHz memory switching rate and 32·10$^{-5}$ for the 50 kHz switching of the memory. In the latter case the error was higher due to the frequency resolution of the DAQ card.

## 5. Conclusions

We present a superconducting kinetic inductance memory ("SKIM element"), which can operate at temperatures as high as 2.8 K, in a cryogen free cooler. The SKIM element is a fully superconducting device, containing no normal-metal dissipative elements. This non-volatile memory device stores information in a form of single flux quanta. Our SKIM was operational up to ~50 kHz, the frequency being limited by the setup. In the future the system can be coupled to a microwave readout system, thus increasing the rate drastically. The SKIM memory shows a very low error rate (~10$^{-5}$), which is much lower than previously reported in nanowire memory elements.

**Acknowledgments**
The work was supported by the Army grant W911NF-18-1-0117.